\documentclass[11pt]{article}
\usepackage{hyperref}
\pdfoutput=1
\begin{document}
\title{Broad Leaves in Strong Flow}
\author{Laura A. Miller and Arvind Santhanakrishnan \\
\\\vspace{6pt} Departments of Mathematics and Biology, \\ University of North Carolina, Chapel Hill, NC 27599, USA \\
\\\vspace{6pt} Department of Mechanical Engineering, \\ Oklahoma State University, Stillwater, OK  74078, USA}
\maketitle
\begin{abstract}
Flexible broad leaves are thought to reconfigure in the wind and water to reduce the drag forces that act upon them. Simple mathematical models of a flexible beam immersed in a two-dimensional flow will also exhibit this behavior. What is less understood is how the mechanical properties of a leaf in a three-dimensional flow will passively allow roll up into a cone shape and reduce both drag and vortex induced oscillations. In this fluid dynamics video, the flows around the leaves are compared with those of simplified sheets using 3D numerical simulations and physical models. For some reconfiguration shapes, large forces and oscillations due to strong vortex shedding are produced. In the actual leaf, a stable recirculation zone is formed within the wake of the reconfigured cone. In physical and numerical models that reconfigure into cones, a similar recirculation zone is observed with both rigid and flexible tethers. These results suggest that the three-dimensional cone structure in addition to flexibility is significant to both the reduction of vortex-induced vibrations and the forces experienced by the leaf.
\end{abstract}
\section{Introduction}

This video shows how leaves and other flexible sheets reconfigure in strong
flow. The motion of the leaves in flow are compared with those of simplified physical and numerical models of flexible sheets
attached to both rigid and flexible beams. In the actual leaf, a stable recirculation zone is formed
within the wake of the reconfigured cone. In physical models that reconfigure into cones, a similar
recirculation zone is observed with both rigid and flexible tethers.

Experiments with physical models show that flexible rectangular sheets that reconfigure into U-shapes
are less stable than cones when attached to flexible tethers. In these cases, larger forces
and oscillations due to strong vortex shedding are observed. These results suggest that the three-dimensional
cone structure in addition to flexibility is significant to both the reduction of vortex-induced
vibrations and the forces experienced by the leaf.

\section{Methods}
The experimental portion of this work was conducted at the University of North Carolina at Chapel Hill, in the facilities of the Joint Applied Math and Marine Science Fluids Lab. Video imaging was used as the primary diagnostic technique to examine the leaf orientation in air moving at speeds of 1-10 meters per second in an open wind tunnel as described by Vogel [3]. Leaves shown in the video include the wild ginger, \textit{Hexastylis arifolia}, the tulip popular, \textit{}, and the maple, \textit{}.

Numerical simulations of the fully-coupled fluid-structure interaction problem were performed using a distributed-memory parallel implementation of the immersed boundary method with support for Cartesian grid adaptive mesh refinement (AMR) [1]. All simulations were performed at a Reynolds number of 250, and the dimensionless force generated as well as the reconfiguration shapes are similar to what occurs for the high Reynolds number flows relevant to broad leaves. More details of these simulations can be found in Miller \textit{et al.} [2].

\section{References}
[1] Griffith, B. E., Hornung, R. D., McQueen, D. M., and Peskin, C. S. An adaptive, formally second order accurate version of the immersed boundary method. \textit{J. Comput. Phys.}, 223(1):10-49, 2007.

[2] Miller, L. A., Santhanakrishnan, A., Jones, S., Hamlet, C. L., Mertens, K., and Zhu L.
Reconfiguration and the reduction of vortex-induced vibrations in broad leaves. \textit{J. Exp. Biol.}
215(15): 2716-2727, 2012.

[3] Vogel, S. Drag and Reconfiguration of Broad Leaves in High Winds, \textit{J. Exp. Botany}, 40, 941-948, 1989.
\end{document}